\documentstyle[multicol,prb,aps]{revtex}

\begin{document}
\title{Theory of electrical spin injection: \\
Tunnel contacts as a solution of the conductivity mismatch problem}
\author{ E. I. Rashba\cite{address} }
\address{ Department of Physics, MIT, Cambridge, Massachusetts 02139  }
%
\date{September 11, 2000}
\maketitle

\begin{abstract}                
Theory of electrical spin injection from a ferromagnetic (FM) metal
into a normal (N) conductor is presented. We show that tunnel
contacts (T) can dramatically increase spin injection and solve the
problem of the mismatch in the conductivities of a FM metal and a
semiconductor microstructure. We also present explicit expressions for
the spin-valve resistance of FM-T-N- and FM-T-N-T-FM-junctions with 
tunnel contacts at the interfaces and show that the resistance includes 
both positive and negative contributions (Kapitza resistance and 
injection conductivity, respectively).
\end{abstract}


\begin{multicols}{2}

{\it Introduction.} Since the seminal proposal by Datta and Das of a spin 
transistor\cite{DD90} based on spin precession controlled by an
external electric field {\it via} spin-orbit (SO) coupling,\cite{R60} 
there exists persistent and growing interest in spin injection into 
semiconductor microstructures. For a spin transistor to work 
(i) long spin relaxation time in a semiconductor, (ii) gate voltage
control of the SO coupling, and (iii) high spin injection
coefficient are needed. Slow relaxation of electron spins in
semiconductors has been established by optical experiments.
\cite{OptOr} Modulation of the SO splitting at the 
Fermi level by gate voltage has been reported for both electrons 
and holes and for different semiconductor materials.
\cite{SHMCSL96,NiLu,EnHeHu,SS99,G00} 
Theory of the gate voltage effect has been developed in many detail.
\cite{LMR88,SRBPZW}

However, as distinct from spin injection from a FM source into a 
paramagnetic metal, very efficient and well documented experimentally,
\cite{sp-met} spin injection from a similar source into a semiconductor
\cite{AP76} remains a challenging task. After numerous efforts, promising 
results have been reported recently.\cite{HBYJ99,GSBLR99,MGMM00}
Unfortunately, spin polarization measured in Refs.~\onlinecite{HBYJ99} 
and \onlinecite{GSBLR99}  was only about 1\%. Problems 
with injection from metallic contacts promoted the idea to use a 
semimagnetic semiconductor as a spin aligner, \cite{simi0} and high 
degree of spin polarisation has been achieved in this way.\cite{semi} 
However, FM metal sources remain an indispensable tool for room 
temperature devices.

Schmidt {\it al.}\cite{SMFW00} revealed that the basic obstacle for spin
injection from a FM metal emitter into a semiconductor originates 
from the conductivity mismatch between these materials. They have shown, 
that in a diffusive regime the spin injection coefficient $\gamma$ is
$\gamma \propto \sigma_N/\sigma_F \ll 1$, where $\sigma_N$ and $\sigma_F$ 
are conductivities of the N (semiconductor) and FM (metallic emitter) 
contacts, respectively. Their result explains, in a natural way, the 
striking difference 
between emission from a FM metal into a paramagnetic metal with 
$\sigma_N/\sigma_F \agt 1$ and a semiconductor with 
$\sigma_N/\sigma_F \ll 1$. At first glance, the problem seems
insurmountable. However, we show in this paper that insertion of a tunnel
contact T at a FM-N interface can remedy it. This contact takes control 
over $\gamma$ and  eliminates the conductivity mismatch. For this 
purpose, tunnel resistance $r_c$ does not need to be really large. 
It should only be larger than competing ``effective resistances'' 
making the total contact resistance: 
\begin{equation}
 r_c \agt L_F/\sigma_F,~ {\rm min}\{L_N, w\}/\sigma_N , 
\label{eq1}
\end{equation}
where $L_F$ and $L_N$ are spin diffusion lengths in the FM and N conductors,
respectively, and $w$ is the N conductor width. 

It is our general conclusion that {\it the spin injection coefficient is
controlled by the element of a FM-T-N-junction having the largest effective 
resistance.}

Since the dependence of the FM-T-N-T-FM-junction resistance 
${\cal R}_{\rm j}$ on the mutual polarisation of FM electrodes
(spin-valve effect) is used for spin injection detection, we have calculated 
${\cal R}_{\rm j}$. It originates from the current conversion in the junction 
and includes, side by side with a positive term (Kapitza resistance), a 
negative term (injection conductivity) originating from spin injection 
and proportional to $\gamma^2$. This term has never appeared in the 
literature before.

{\it Theory of a FM-T-N-junction}. To make the effect of a 
tunnel contact most clear, we simplify the problem of a FM-T-N-junction 
 between semiinfinite FM $(x<0)$ and N $(x>0)$ conductors as much as 
possible. We apply the diffusion approximation and suppose that the 
T contact, at $x=0$, is spin selective, i.e., has different 
conductivities, $\Sigma_{\uparrow}$ and $\Sigma_{\downarrow}$, for 
up and down spins, respectively, and there is no spin relaxation in it. 
Therefore, the problem differs from that considered by van Son 
{\it et al.}.\cite{vanSon} only by the presence of the T contact. 
Because of some subtleties in calculating the potential distribution 
near spin emitting contacts, we outline the procedure in some detail.

In the approximation linear in the total current $J$, the currents  
$j_{\uparrow,\downarrow}(x)$ carried by up- and down-spins can be
written in terms of the space derivatives of electrochemical potentials 
$\zeta_{\uparrow,\downarrow}(x)$,
\begin{equation}
 j_{\uparrow,\downarrow}(x)=\sigma_{\uparrow,\downarrow}
\zeta^{~\prime}_{\uparrow,\downarrow}(x),  
\label{eq2}
\end{equation}
which are related to the non-equilibrium parts $n_{\uparrow,\downarrow}(x)$ 
of the electron concentrations  and to the electrical potential 
$\varphi_F(x)$ in the FM region by equations
\begin{equation}
 \zeta_{\uparrow,\downarrow}(x)= (eD_{\uparrow,\downarrow}/
\sigma_{\uparrow,\downarrow})n_{\uparrow,\downarrow}(x) - \varphi_F(x),
\label{eq3}
\end{equation}
and $D_{\uparrow,\downarrow}$ and $\sigma_{\uparrow,\downarrow}$ are
diffusion coefficients and conductivities, respectively, of up- and down-spin 
electrons. These equations should be supplemented by the equation
\begin{equation}
  n_{\uparrow}(x) + n_{\downarrow}(x) =0,
\label{eq4}
\end{equation}
maintaining the electrical neutrality under the spin injection conditions, 
and the continuity and  charge conservation equations
\begin{equation}
   j^{~\prime}_{\uparrow}(x) = en_{\uparrow}(x)/\tau^F_s,~
 J = j_{\uparrow}(x)+j_{\downarrow}(x)= {\rm const},
\label{eq5}
\end{equation}
where $\tau^F_s$ is the spin relaxation time. In the ``metallic''
approximation, Eq.~(\ref{eq4}) is equivalent to a Poisson equation and 
connects transport in the both spin channels.

Let us introduce symmetric in spins variables
\begin{equation}
  \zeta_F(x)= \zeta_{\uparrow}(x)- \zeta_{\downarrow}(x),~
j_F(x)=j_{\uparrow}(x)-j_{\downarrow}(x).
\label{eq6}
\end{equation}
In these notations, the standard routine results in a diffusion equation
\begin{equation}
  D_F \zeta^{~\prime \prime}_F(x)=  \zeta_F(x)/\tau^{F}_s,~
D_F=(\sigma_{\downarrow}D_{\uparrow}+\sigma_{\uparrow}D_{\downarrow})/
\sigma_F, 
\label{eq7}
\end{equation}
where $\sigma_F=\sigma_{\uparrow}+\sigma_{\downarrow}$.
 The equation for $\varphi_F(x)$,
\begin{equation}
  \varphi^{~\prime}_F(x)=[(D_{\uparrow}-D_{\downarrow})/D_F]
(\sigma_{\uparrow}\sigma_{\downarrow}/{\sigma_F^2})\zeta^{~\prime}_F(x) 
- J/{\sigma_F},
\label{eq8}
\end{equation}
also follows from (\ref{eq2})--(\ref{eq5}). Restricting ourselves 
with zero temperature, $T=0$,\cite{temp} it is convenient to introduce 
densities of states at the Fermi level, $\rho_{\uparrow,\downarrow}$, and to 
apply Einstein relations $e^2D_{\uparrow,\downarrow}=
\sigma_{\uparrow,\downarrow}/\rho_{\uparrow,\downarrow}$. The identities
\begin{eqnarray}
  e^2 D_{F}&=&(\sigma_{\uparrow}\sigma_{\downarrow}/\sigma_F )
({\rho_F}/{\rho_{\uparrow}\rho_{\downarrow}}), \nonumber \\ 
(\rho_{\downarrow}\sigma_{\uparrow}-\rho_{\uparrow}\sigma_{\downarrow})&/&
{\rho_F \sigma_F}=
[({\Delta\sigma}/{\sigma_F})-({\Delta\rho}/{\rho_F})]/2, 
\label{eq9}
\end{eqnarray}
where $\Delta\sigma=\sigma_{\uparrow}-\sigma_{\downarrow}$, 
$\Delta\rho=\rho_{\uparrow}-\rho_{\downarrow}$,  
$\rho_F = \rho_{\uparrow}+\rho_{\downarrow}$, allow us to rewrite 
Eq.~(\ref{eq8}) as
\begin{equation}
  \varphi^{~\prime}_F(x)=[(\Delta\sigma/\sigma_F) -
({\Delta\rho}/{\rho_F})]\zeta_F^{~\prime}(x)/2-J/{\sigma_F}.
\label{eq10}
\end{equation}
It follows from (\ref{eq2}) and (\ref{eq10}) that:
\begin{equation}
 j_F(x)= 2~(\sigma_{\uparrow}\sigma_{\downarrow}/\sigma_F)
\zeta^{~\prime}_F(x)+({\Delta\sigma}/{\sigma_F})J.
\label{eq11}
\end{equation}

Eqs.~(\ref{eq7}), (\ref{eq10}), and (\ref{eq11}) make a complete system
of bulk equations for the F region. They also determine  
\begin{equation}
\zeta_{\uparrow}(x)+\zeta_{\downarrow}(x)=-[2\varphi_F(x) +
(\Delta\rho/\rho_F)\zeta_F(x)]
\label{eq11a}
\end{equation}
and $n_{\uparrow}(x)=
(\rho_{\uparrow}\rho_{\downarrow}/{\rho_F})\zeta_F(x)$.
Equations for the N region can be obtained from them by putting 
$\sigma_{\uparrow}=\sigma_{\downarrow}=\sigma_{N}/2,~ 
\Delta\rho=\Delta\sigma=0$, and $D_N=D_{\uparrow}=D_{\downarrow}$:
\begin{eqnarray}
 D_N \zeta^{~\prime \prime}_N(x)&&=  \zeta_N(x)/\tau^N_s,
~\varphi^{~\prime}_N(x)=-J/\sigma_N,~\nonumber \\
&&j_N(x)=\sigma_N \zeta^{~\prime}_N(x)/2.
\label{eq12}
\end{eqnarray}

The boundary conditions at $x=0$ follow from the absence of spin 
relaxation at the interface. The current $j_{\uparrow}(x)$ is continuous 
at $x=0$ and the condition  $j_F(0)=j_N(0)$, according (\ref{eq11}) 
and (\ref{eq12}), can be rewritten as
\begin{equation}
  \sigma_N \zeta^{~\prime}_N(0) - 
4(\sigma_{\uparrow}\sigma_{\downarrow}/{\sigma_F})\zeta^{~\prime}_F(0)=
2(\Delta\sigma /{\sigma_F})J.
\label{eq13}
\end{equation}
Here $\zeta^{~\prime}_F(0)$ and $\zeta^{~\prime}_N(0)$ are the values of 
$\zeta^{~\prime}(x)$ at the left and the right sides of the interface, 
respectively. Low tunnel transparency of the contact supports differences 
in the potentials $\zeta^F_{\uparrow,\downarrow}$ and 
$\zeta^N_{\uparrow,\downarrow}$ at the F and N sides of it, and makes
$\zeta_{\uparrow, \downarrow}(x)$ discontinuous at $x=0$. \cite{tau} Similar 
to (\ref{eq2}), these differences are related to the currents as
\begin{equation}
 j_{\uparrow,\downarrow}(0)=\Sigma_{\uparrow,\downarrow}
(\zeta^N_{\uparrow,\downarrow}-\zeta^F_{\uparrow,\downarrow}),
\label{eq14} 
\end{equation}
or, in the symmetric variables of Eq.~(\ref{eq6}), as
 \begin{equation}
\zeta_N(0)-\zeta_F(0)=-2~({\Delta\Sigma}/\Sigma)~r_c J+2r_c j(0),
\label{eq15}  
\end{equation}
\begin{equation}
  (\varphi_F(0)-\varphi_N(0))+{{\Delta\rho}\over{2\rho_F}}~\zeta_F(0)=
r_cJ-{{\Delta\Sigma}\over{\Sigma}}~r_cj(0),
\label{eq16}
\end{equation}
where Eq.~(\ref{eq11a}) has been taken into account. The current 
$j(0)=j_F(0)=j_N(0)$ should be found from (\ref{eq11}) or (\ref{eq12}).
Here $\Delta\Sigma=\Sigma_{\uparrow}-\Sigma_{\downarrow}$, 
$\Sigma=\Sigma_{\uparrow}+\Sigma_{\downarrow}$, and 
$r_c=\Sigma/4\Sigma_{\uparrow}\Sigma_{\downarrow}$ is the effective
contact resistance.   

One important conclusion follows from (\ref{eq16}) immediately: a finite 
voltage drop at the interface, 
$V_{\rm if}=\varphi_F(0)-\varphi_N(0)\propto J$, exists even for
 $r_c=0$ because of $\Delta\rho\neq 0$. This fact is not surprising.  
Similar discontinuities exist at abrupt 
{\it  p-n}-junctions\cite{S49} and near current converting surfaces 
in the theory of the diffusion size effect.\cite{KR} They should also
contribute to the giant magnetoresistance.\cite{P98}

{\it Injection coefficient.} Solutions of (\ref{eq7}) and 
(\ref{eq12}) for $\zeta_{F,N}(x)$ are exponents decaying with the 
diffusion lengths $L_F=(D_F \tau_s^F)^{1/2}$ and
$L_N=(D_N\tau_s^N)^{1/2}$. 
Therefore, 
$\zeta^{\prime}_N(0)=-\zeta_N(0)/L_N=2\gamma J/{\sigma_N}$ and 
$\zeta^{\prime}_F(0)=\zeta_F(0)/L_F$. Let us define the injection 
coefficient as $\gamma=j(0)/J$. Eliminating $\zeta_F(0)$ from (\ref{eq13}) 
and (\ref{eq15}), we get
\begin{equation}
 \gamma = [r_F~({\Delta\sigma}/{\sigma_F})+
r_c~({\Delta\Sigma}/{\Sigma})]/r_{FN},
\label{eq19} 
\end{equation}
where $r_{FN} = r_F+r_N+r_c$, 
$r_F=L_F\sigma_F/4\sigma_{\uparrow}\sigma_{\downarrow}$, and 
$r_N=L_N/\sigma_N$. The equation for $r_{FN}$ shows that $r_c$, 
 $r_F$ and $r_N$ are connected in series.
It follows from (\ref{eq19}) that with $r_F \ll r_N$, the injection
coefficient can be large, $\gamma \sim 1$, if and only if $r_c \agt r_N$, 
in agreement with (\ref{eq1}). This criterion is rather soft and
is satisfied for narrow tunnel junctions of the atomic scale. Actually,  
any kind of a spin selective contact with high resistance $r_c$ suits
this criterion. For $r_c\gg r_N, r_F$, the injection coefficient 
$\gamma\approx {\Delta\Sigma}/{\Sigma_F}$.  In this regime the contact 
takes control over $\gamma$ and completely determines it.

{\it Spin-e.m.f.} The same FM-T-N-junction can be used for detecting spin 
accumulation $n_{\infty}$ homogeneously produced in the N region by some 
external source by measuring open circuit voltage (floating potential) on a FM
electrode. This signal is some kind of photo-e.m.f. and has been
successfully used by Johnson\cite{J93} for detecting spins injected into
paramagnetic metals, while absence of a similar signal from semiconductor
heterostructures\cite{FHJWDB00} signifies low spin injection level. 
Derivations similar to the presented above result in a spin-e.m.f. signal
\begin{equation}
  \varphi_F = 2\gamma (eD_N/L_N)r_N n_{\infty},
\label{eq20}
\end{equation}
where $\gamma$ is defined by (\ref{eq19}), and 
$ n_{\infty}=n_{\uparrow}(x=\infty)$. Large value of $r_N$, while
suppressing spin injection, facilitates large magnitude of $\varphi_F$.

{\it Resistance of a FM-T-N-junction.} The voltage drop at the interface, 
$V_{\rm if}$, permits one to define the interface resistance 
$R_{\rm if}=V_{\rm if}/J$. Finding $\zeta_F(0)$ from (\ref{eq15}) and 
substituting it into (\ref{eq16}), one gets after some algebra:
  \begin{eqnarray}
  R_{\rm if}(\gamma)&=&\Sigma^{-1}+
[r_F({\Delta\rho}/\rho_F)({\Delta\sigma}/\sigma_F)+
r_c ({\Delta\Sigma}/\Sigma)^2]\nonumber \\
&-& \gamma
[r_F({\Delta\rho}/\rho_F)+ r_c({\Delta\Sigma}/\Sigma)].
\label{eq21}
  \end{eqnarray}
The first term in (\ref{eq21}) is an intrinsic property of the interface 
and does not
depend on the presence of non-equilibrium spins, while the last two terms 
cancel when $L_N, L_F \rightarrow 0$. $R_{\rm  if}-{\Sigma}^{-1}$ 
is usually positive but under some conditions, e.g., $\Delta\Sigma =0,~ 
\Delta\sigma/{\Delta\rho} <0$, it is negative.

In addition to $V_{\rm if}$, there exists a potential drop in the regions 
about $L_F$ and $L_N$ around the interface which is of the same order of 
magnitude as $R_{\rm if}-{\Sigma}^{-1}$. The total resistance of the 
junction $R_{\rm j}$ can be found by integrating Eqs.~(\ref{eq8}) 
and (\ref{eq12}) for $\varphi_F$ and $\varphi_N$ and finding the 
integration constant from  (\ref{eq16}). Subtracting the voltage 
drop over the nominal resistances of the FM and N regions
from the potential difference between their ends, we get
\begin{eqnarray}
  R_{\rm j}(\gamma, r_{FN}) &=& \Sigma^{-1} \nonumber \\
 &+& [r_F ({\Delta\sigma}/{\sigma_F})^2 +
r_c ({\Delta\Sigma}/\Sigma)^2]-{\gamma}^2r_{FN}.
\label{eq22}
\end{eqnarray}
Two last terms in (\ref{eq22}) originate from non-equilibrium spins
and cancel when $L_N, L_F \rightarrow 0$. The second term in 
(\ref{eq22}) is positive and can be identified as {\it Kapitza resistance} 
originating from the conversion of spin flows. The third term is negative 
and explicitly related to the spin injection. Therefore, we term it 
{\it injection conductivity}. The sum of both non-equilibrium terms in
(\ref{eq22}) is always positive. It is interesting to note that the factor 
${\Delta\rho}/{\rho_F}$ which is present in $R_{\rm if}$ cancels from
$R_{\rm j}$. Resistances $R_{\rm if}$ and $R_{\rm j}$ can be measured
separately in spin-e.m.f. and spin-valve experiments. For $r_c=0$, 
Eqs.~(\ref{eq19}) and (\ref{eq22}) are equivalent to the results by van 
Son {\it et al.}\cite{vanSon}, hence, the resistance found by them should 
be identified as $R_{\rm j}$. 

{\it Spin injection into a FM-T-N-T-FM-junction}. General equations derived
above are also applicable to a system with two interfaces, two tunnel 
contacts, and an N region between them. We attach indices 
L and R to the parameters of the left and right ferromagnets and tunnel 
contacts and neglect spin relaxation in N region since in this case
equations simplify and the problem can be solved in terms of the
parameters of a single FM-T-N-junction. Writing equations similar to 
(\ref{eq13}) and (\ref{eq15}) for each contact, taking their sums, and 
eliminating $\zeta^R_F-\zeta^L_F$ ~($\zeta^R_F$ and $\zeta^L_F$ being 
values of $\zeta_F(x)$ at the junction boundaries), one finds injection
coefficient $\Gamma= (j_{\uparrow}^N-j_{\downarrow}^N)/J$:
\begin{equation}
\Gamma= (r^L_{FN}\gamma_L + r^R_{NF}\gamma_R)/r_{FNF},
\label{eq23}
\end{equation}
where $r_{FNF} = r_N^w + r^L_F + r^R_F + r_c^L + r_c^R$,
$r_N^w=w/\sigma_N$ is a nominal resistance of the N region, $w$ is 
its width, and $\gamma_L$ and $\gamma_R$ can be found from Eq.~(\ref{eq19}) 
for L and R interfaces. Similar to (\ref{eq19}),  injection is controlled 
by the larger of the resistances $r_F^{L,R}$ and $r_c^{L,R}$. To achieve a 
large $\gamma$ value it is enough to have only one tunnel contact, either the 
left or the right one. The second contact is only needed for detecting spin 
injection by the spin-valve effect. Even in a completely antisymmetric system, 
${\Delta\sigma_L}/{\sigma_L}=-{\Delta\sigma_R}/{\sigma_R},
~ {\Delta\Sigma_L}/{\Sigma_L} =-{\Delta\Sigma_R}/{\Sigma_R},~r_c^L=r_c^R$, 
non-equilibrium spins are present in the N region. E.g., for
$r_c^L=r_c^R=0$, their concentration equals 
$n^N_{\uparrow}(x)=-(\sigma_N/2D_N)({{\Delta\sigma}_L/\sigma_F})r_F J$ 
= const, and the result $\Gamma =0$ following from (\ref{eq23}) is 
tantamount to the absence of diffusion currents in the N-region. 
Non-equilibrium spins in it can be detected by spin-e.m.f. 

{\it Resistance of a FM-T-N-T-FM-junction}. Similar to a FM-N-junction,
 $\varphi(x)$ shows abrupt change at both interfaces and gradual 
change near them at the scale of $L_F$. Interfacial resistances 
${\cal R}_{\rm if}^{L,R}$ are similar to 
(\ref{eq21}):
\begin{equation}
 {\cal R}_{\rm if}^{L,R} = R_{\rm if}^{L,R}(\Gamma, r_{FNF}),
\label{eq25}
\end{equation}
i.e., they can be found from (\ref{eq22}), however, with $\Gamma$ instead 
of $\gamma$ and $r_{FNF}$ instead of $r_{FM}$. The junction resistance 
${\cal R}_{\rm j}$ can be written in a similar way in terms of $R_{\rm j}$:
\begin{equation}
  {\cal R}_{\rm j} = r_N^w +  
R_{\rm j}^L(\Gamma, r_{FNF}) + 
R_{\rm j}^R(\Gamma, r_{FNF}),
\label{eq26}
\end{equation}
i.e., it can be found from Eq.~(\ref{eq22}) by plugging into it the
parameters of both contacts and changing $\gamma\rightarrow\Gamma$ and
$r_{FN}\rightarrow r_{FNF}$. Therefore, ${\cal R}_{\rm j}$ also includes 
the Kapitza resistance and injection conductivity. The non-equilibrium 
part of ${\cal R}_j$ is always positive, but the explicit equation
proving this fact is somewhat lengthy.   

Let us mention that Eqs.~(\ref{eq25}) and (\ref{eq26}) for resistances, 
as well as Eqs.~(\ref{eq21}) and (\ref{eq22}), include only products or 
squares of the differences $\Delta\sigma$, $\Delta\Sigma$ and $\Delta\rho$, 
while the equations for potentials [like Eqs.~(\ref{eq13}), (\ref{eq15}) and  
(\ref{eq16})] include them in the first power. 

Detection of spin injection by the spin-valve effect is based on the 
change in  ${\cal R}_{\rm j}$ when the magnetization direction of one of 
the two identical FM-electrodes is reversed
($\Delta \sigma_R \rightarrow - \Delta \sigma_R,~
\Delta \Sigma_R \rightarrow - \Delta \Sigma_R$). It comes exclusively
from the injection conductivity, 
$\Delta {\cal R}_{\rm j} = r_{FNF}({\Gamma}^{~2}_{\uparrow \uparrow}- 
{\Gamma}^{~2}_{\uparrow \downarrow})$, and equals:
\begin{equation}
\Delta {\cal R}_{\rm j} = \gamma_L \gamma_R (4r^L_{FN}r^R_{NF}/r_{FNF}).
\label{eq27}
\end{equation}

The resistance ${\cal R}_{\rm j}$, 
Eq.~(\ref{eq26}), remains finite for a completely antisymmetric 
system even for $|\Delta\sigma|\ll \sigma_F$. In the absence of spin
relaxation in the N region and at both contacts, no conductivity through a
junction could be expected at the first glance. However, it exists and
its mechanism is as follows. Because $\Gamma_{\uparrow \downarrow}=0$, 
the currents $j^N_{\uparrow}$ and $j^N_{\downarrow}$ in the N region 
are driven only by the electric field and are equal exactly, 
$j^N_{\uparrow}= j^N_{\downarrow}= J/2$. In the FM regions, the
currents of the minority spins are driven mostly by
diffusion. Therefore, the concentrations of non-equilibrium spins near
the interfaces equal $n \approx (L_F/2eD_F)J$, and the
only restriction on the diffusion current comes from the condition
that the total concentration of minority carriers is positive, i.e., 
$n^0_{\rm min}\pm n > 0$, where $n^0_{\rm min}$ is the equilibrium
concentration of minority carriers. Hence, the Ohmic conductivity of a
FM-T-N-T-FM-junction remains finite even when
$|\Delta\sigma|/\sigma_F \rightarrow 0$, but the Ohmic region 
 becomes  narrower and disappears completely for 
$|\Delta\sigma|=\sigma_F$. Non-linear conductivity is outside the
scope of this paper. 

For $r_c^L=r_c^R=0$, Eq.~(\ref{eq23}) for $\Gamma$ is equivalent to
the result by Schmidt {\it et al}.\cite{SMFW00} However, Eqs.~(\ref{eq26}) 
and (\ref{eq27}) differ from the equations for Ohmic resistance of 
 Refs.~\onlinecite{SMFW00} and \onlinecite{FHJWDB00}.  
Because the derivation procedure has not been specified there, the origin 
of the discrepancy is unclear.

{\it Discussion.} The above theory suggests that tunnel contacts obeying 
criterion (\ref{eq1}) should provide a tremendous increase in spin 
polarization of the currents injected electrically from a FM metal 
into a semiconductor. Our conclusion is based on the assumption that 
spin conductivity ratio $\Delta\Sigma/\Sigma$ is large for tunnel 
contacts. In fact, Alvorado\cite{A95} has shown that  spin
polarization is large for narrow barriers and can reach about
50\%. Different types of tunnel contacts have been successfully used, 
 e.g., STM tips in vacuum\cite{A95} and in air,\cite{Hill}
Schottky barriers,\cite{MVL98,DBTH00} and resonant double barriers.\cite{O98}  
Therefore, inclusion of appropriate barriers into a circuit should be a
soluble problem. One should also bear in mind, that it is not the ballistic
transport, but the ability of tunnel contacts to support a considerable 
difference in electrochemical potentials under the conditions of slow spin
relaxation, which is important for efficient spin injection. Therefore, 
different contacts combining small spin diffusivity with low spin relaxation
rate should possess similar properties.  

In conclusion, we (i) have shown that tunnel contacts can solve 
the problem of the electrical spin injection from a ferromagnetic metal into 
a semiconductor, and (ii) have derived explicit expressions for the 
spin injection coefficient, spin-valve effect, and spin-e.m.f.

I am grateful to Dr. Alexander Efros for useful discussions.

\end{multicols}

\begin{references}
\bibitem[*]{address} e-mail: rashba-step@mediaone.net (E. I. Rashba)
\bibitem{DD90} S. Datta and B. Das, Appl. Phys. Lett. {\bf 56}, 665 (1990).
\bibitem{R60} E. I. Rashba, Sov. Phys. - Solid State {\bf 2}, 1109
  (1960); E. I. Rashba and V. I. Sheka, in: {\it Landau Level
    Spectroscopy}, ed. by G. Landwehr and E. I. Rashba
  (North-Holland, Amsterdam, 1991), v. 1, p. 131.
\bibitem{OptOr} {\it Optical Orientation}, ed. by F. Meier and
  B. P. Zakharchenya (North-Holland, Amsterdam 1984);
 J. M. Kikkawa and D. D. Awschalom,
  Phys. Rev. Lett. {\bf 80}, 4313 (1998);
 D. H\"{a}gele {\it et al.}, Appl. Phys. Lett. {\bf 73}, 1580 (1998).
\bibitem{SHMCSL96} M. Schultz {\it et al.}, Semicond. Sci. Technol. 
{\bf 11}, 1168 (1996).
\bibitem{NiLu} J. Nitta {\it et al.}, Phys. Rev. Lett. {\bf 78}, 1335 (1997);
J. P. Lu {\it et al.}, {\it ibid.} {\bf 81}, 1282 (1998).
\bibitem{EnHeHu} G. Engels {\it et al.}, Phys. Rev. B {\bf 55}, R1958 (1997);
 J. P. Heida {\it et al.}, {\it ibid.} {\bf 57}, 11 911 (1998);
 C.-M. Hu {\it et al.}, {\it ibid.} {\bf 60}, 7736 (1999).
\bibitem{SS99} Y. Sato {\it et al.}, Physica B {\bf 272}, 114 (1999);
 S. Sasa {\it et al.}, {\it ibid.} B {\bf 272}, 149 (1999).
\bibitem{G00} D. Grundler, Phys. Rev. Lett. {\bf 84}, 6074 (2000).
\bibitem{LMR88} G. Lommer, F. Malcher, and U. R\"{o}ssler,
  Phys. Rev. Lett. {\bf 60}, 728 (1988); P. V. Santos and M. Cardona,
  {\it ibid.} {\bf 72}, 432 (1994).
\bibitem{SRBPZW} E. A. de Andrada e Silva, G. C. La Rocca, and
  F. Bassani, Phys. Rev. B {\bf 55}, 16 293 (1997);
 P. Pfeffer and W. Zawadzki, {\it ibid.} {\bf 59}, R5312 (1999);
 R. Winkler, {\it ibid.} {\bf 62}, 4245 (2000).
\bibitem{sp-met} P. M. Todrow and M. Meservey, Phys. Rev. Lett. {\bf
    26}, 192 (1971); M. Johnson and R. H. Silsbee, {\it ibid.} {\bf
    55}, 1790 (1985).
\bibitem{AP76} A. G. Aronov and G. E. Pikus, Sov. Phys. Semicond. {\bf
    10}, 698 (1976).
\bibitem{HBYJ99} P. Hammer {\it et al.}, Phys. Rev. Lett. {\bf 83}, 203 (1999).
\bibitem{GSBLR99} S. Gardelis {\it et al.}, Phys. Rev. B {\bf 60}, 7764
  (1999).
\bibitem{MGMM00} G. Meier, D. Grundler, T. Matsuyama, and U. Merkt, {\it
      Abstracts of the Symp. on Spin-Electronics} (Halle, July 2000). 
\bibitem{simi0} M. Oestreich {\it et al.}, Appl. Phys. Lett. 
{\bf 74}, 1251 (1999).
\bibitem{semi} R. Fiederling {\it et al.}, Nature {\bf 402}, 787 (1999);
  Y. Ohno {\it et al.}, {\it ibid.} {\bf 402}, 790 (1999); 
B. T. Jonker {\it et al.}, Phys. Rev. B {\bf 62}, R4790 (2000). 
\bibitem{SMFW00} G. Schmidt, D. Ferrand, L. W. Mollenkamp, A. T. Filip, 
B. J. van Wees, Phys. Rev. B {\bf 62}, R4790 (2000).
\bibitem{vanSon} P. C. van Son, H. Kempen, and P. Wyder, 
Phys. Rev. Lett. {\bf 58}, 227 (1987).
\bibitem{temp} This restriction does not influence the basic results. 
\bibitem{tau} Criterion (\ref{eq1}) is equivalent to
  $t\alt(\tau_p/\tau_s)^{1/2}$, where $t$ is tunnel transparency, and
  $\tau_p$ ans $\tau_s$ are momentum and spin relaxation times. It
 ensures  discontinuity of $\zeta(x)$ at $x=0$.
\bibitem{S49} W. Shockley, Bell Syst. Tech. J. {\bf 28}, 435 (1949).
\bibitem{KR} V. Ya. Kravchenko and E. I. Rashba, Sov. Phys. - JETP
  {\bf 29}, 918 (1969).
\bibitem{P98} G. A. Prinz, Science {\bf 282}, 1660 (1998).
\bibitem{J93} M. Johnson, Phys. Rev. Lett. {\bf 70}, 2142 (1993).
\bibitem{FHJWDB00} A. T. Filip {\it et al.}, Phys. Rev. B {\bf 62},
  9996 (2000).
\bibitem{A95} S. F. Alvorado, Phys. Phys. Lett. {\bf 75}, 513 (1995).
\bibitem{Hill} C. J. Hill {\it et al.}, cond-mat/0010058 (4 October 2000).
\bibitem{MVL98} D. J. Monsma {\it et al.}, Science
  {\bf 281}, 407 (1998).
\bibitem{DBTH00} V. Delmouly, A. Bournel, G. Tremblay, and P. Hesto,
{\it Abstracts of the Symp. on Spin-Electronics} (Halle, 
    July 2000). 
\bibitem{O98} H. Ohno, Science {\bf 281}, 951 (1998).
\end{references}
\end{document}